\documentstyle[12pt]{article}
\def\laq{\raise 0.4 ex \hbox{$<$}\kern -0.8 em\lower 0.62 ex\hbox{$\sim$}}
\def\gaq{\raise 0.4 ex \hbox{$>$}\kern -0.7 em\lower 0.62 ex\hbox{$\sim$}}

\input{epsf}
\def\PR{{\it Phys. Rev. } }
\def\IJMP{{\it Int. J. Mod. Phys.} }
\def\IC{{\it Icarus }}
\def\AJ{{\it Ap. J. } }
\def\PL{{\it Phys. Lett. }}
\def\CQG{{\it Class. Quant. Gravity }}
\def\JGR{{\it J. Geophys. Res. }}

\begin{document}
\begin{flushright}
{DF/IST-4.2005} \\
{June 2006} \\
\end{flushright}
\vglue 1cm

\begin{center}
{{\bf  Pioneer Anomaly and the Kuiper Belt mass distribution}}

\vglue 0.5cm
{O.\ Bertolami and P. Vieira}

\vglue 0.2cm

{E-mail addresses: {\tt orfeu@cosmos.ist.utl.pt; paula.vv@gmail.com}}

\bigskip
{\it Instituto Superior T\'ecnico,
Departamento de F\'\i sica,\\}
\smallskip
{\it Av.\ Rovisco Pais, 1049-001 Lisboa, Portugal\\}
\end{center}

\setlength{\baselineskip}{0.7cm}

\vglue 0.7cm

\centerline{{\bf Abstract}}

\vglue 0.7cm

\noindent

Pioneer 10 and 11 were the first probes sent to study the outer planets of the Solar System and Pioneer 10 was 
the first spacecraft to leave the Solar System.
Besides their already epic journeys, Pioneer 10 and 11 spacecraft were subjected to 
an unaccounted effect interpreted as a constant acceleration toward the Sun,
the so-called Pioneer anomaly.
One of the possibilities put forward for explaining the Pioneer anomaly is the gravitational acceleration of the Kuiper Belt. 
In this work we examine 
this hypothesis for various models for the Kuiper Belt mass distribution. We
find that the gravitational effect due to the Kuiper Belt cannot account for the Pioneer anomaly.
Furthermore, we have also studied the hypothesis that drag forces can explain the Pioneer Anomaly, however we conclude that the
density required for producing the Pioneer anomaly is many orders of 
magnitudes greater than the ones of interplanetary and interstellar dust.

Our conclusions suggest that only through a mission, the Pioneer anomaly can be confirmed and further investigated. If a 
mission with these aims is ever sent to space, it turns out, on account of our results, that it will be also a quite 
interesting probe to study the mass distribution of the Kuiper Belt.

\vfill

\noindent

\vglue 0.7cm

\pagestyle{plain}

\setcounter{equation}{0}
\setlength{\baselineskip}{0.7cm}

\newpage
\section{Introduction}

Pioneer 10 and 11 were launched in 1972 and 1973 to study the outer planets of the Solar System. 
Both probes have followed hyperbolic trajectories close to 
the ecliptic to opposite outward directions in the Solar System \cite{Anderson1}. 
Due to their simple and robust design it was possible to determine their position 
with great accuracy.
During the first years of its life, the acceleration caused by solar radiation pressure  on the Pioneer 10 was 
the main effect \cite{Anderson1}. At about 20 AU (by early 1980s) that effect became sub-dominant and it was 
possible to identify an unaccounted effect. 
This anomaly can be interpreted as a constant acceleration with a magnitude of 
$a=(8.74 \pm 1.33) \times 10^{-10}~m s^{-2}$ and is directed toward the Sun. 
This effect became known as Pioneer anomaly. For the Pioneer spacecraft, it 
has been observed, at least, until $70$ AU \cite{Anderson1}. It was also observed in Pioneer 11 \cite{Anderson1,Bertolami0}.     

We mention that the effect of the Pioneer anomaly on comets, 
asteroids and the outer planets has been discussed in Refs. \cite{Whitmire,Page} and found that data is so far inconclusive. 

One of the discussed possibilities for explaining this anomaly is the gravitational effect of the Kuiper Belt \cite{Anderson1}. 
The subject has been recently discussed \cite{Diego}, and claimed that if one assumes some fairly unrealistic assumptions 
for the Kuiper Belt structure, namely that it starts at about $20$ AU and has mass total mass that is about ten times greater than the 
usually accepted one, than the Pioneer anomaly can be explained. However, as we shall see, 
our conclusions do not support this claim. In this work we examine
this hypothesis through the study of various possible models for the mass distribution of the Kuiper Belt and show that in any of 
them one does not manage to fit the constant deceleration felt by the Pioneer spacecraft as well as the order of magnitude of 
the observed acceleration. 

We also investigate the amount of matter required to explain the anomalous acceleration via the action of drag forces and find that, 
likewise 
for the case of the Kuiper Belt, there is a considerable mismatch of orders of magnitude.

Before closing this introduction it is relevant to point out that the Pioneer anomaly requires for sure, 
some further confirmation, even though, mechanical and 
engineering causes seem to be ruled out (see however Ref. \cite{Scheffer}). This means that evidence points toward new 
physics causes. The most promising new physics explanations for the anomaly include a scalar field with a 
suitable potential \cite{Bertolami1}, the 
running effect of gravitational coupling \cite{Bertolami2, Reynaud}, etc. A fairly complete list of theoretical proposals 
can be found in Ref. \cite{Bertolami1}.

This papers is organized as follows: In section 2 we discuss the main features of the Kuiper Belt structure, 
its mass distribution and the resulting gravitational acceleration. In
section 3 we consider the effect of drag forces on the spacecraft and quantify the densities required to account for 
the anomaly. Finally in section 4, we present our conclusions.


\section{Kuiper's Belt Gravitational Force and other effects}

In this work we test the gravitational effect of four different models for the Kuiper Belt mass distribution: 
two-ring, uniform disk, non-uniform disk and uniform torus models.
The non-uniform disk was considered in the pioneering work of Boss and Peale \cite{Boss}. The two-ring and the uniform disk models 
were considered in Ref. \cite{Anderson1}.
The total mass of Kuiper Belt is taken to $0.3$ Earth masses (this is shown not to qualitatively affect our conclusions).
This is the maximum value allowed for the Kuiper Belt dust 
obtained from far-infrared emission \cite{Backman}. For an extensive discussion on the Kuiper Belt see, for instance, Ref. 
\cite{Lacerda}. 

Before examining in detail the effect of the gravitational 
field of the Kuiper Belt one may wonder whether other effects such as, for instance, 
the migration of debris can be at the origin of the Pioneer anomaly. 
However, evidence points otherwise. Indeed, Pluto's orbit is highly eccentric ($e=0.25$), inclined ($i=17^{o}$) and lies in a 
$3:2$ resonant orbit with Neptune.
Pluto and plutinos have a mechanism that prevents close encounters with Neptune.   
The relative motion of Neptune and Pluto 
ensures that when Pluto is at a crossing point of Neptune's orbit they are far away. 
The minimum distance between them is $17$ AU.
Pluto has a libration of the argument of the perihelion, this puts the perihelion at almost 
the maximum distance away from the ecliptic, which maintains the orbital stability \cite{Malhotrasite,kbo,Malhotra}.
Most likely, Pluto was formed in a heliocentric orbit in the solar nebula instead of
the planetary disk. Malhotra \cite{Malhotra} suggests that Pluto was formed in a heliocentric orbit, circular and with 
low inclination with respect to Neptune. Also, it is 
possible that Pluto and other trans-Neptunian objects have been caught in resonance orbits with Neptune 
as a result of the formation of the outer Solar System. 

In the late stages of planet formation, planets and, in particular the giant ones, were surrounded by planetesimal's debris.
Subsequently, this debris was removed by the gravity of the planets, thereby causing the evolution on their orbits (via 
energy and angular momentum exchange) and the formation of the Oort Cloud. 
The numerical simulations performed in Ref. 
\cite{Fernandez} show that Jupiter's orbit migrated inward, in opposition to the other giant planets orbits which migrated outward.
Thanks to Neptune's outward migration, it has been possible to capture Pluto and others 
trans-Neptunian objects. A consequence of this evolution is an increase in the eccentricity and, in some cases, also an increase 
in the inclination of the orbit of the captured object.  
A fairly general conclusion is that, migration effects imply, for objects beyond Jupiter, in a pressure outward 
the Solar System and hence the Pioneer anomaly cannot be accounted by effects of this nature.
    

\subsection{Mass distribution and gravitational acceleration}

Let us now consider in detail the gravitational forces
generated by the various models of the Kuiper Belt. The models were
chosen to be consistent with the main observational features of the Kuiper Belt. The total mass
of the Kuiper Belt was considered to be the same for all the models and
taken to be about $0.3$ Earth masses, a figure that is often found in the literature \cite{Backman}.

Kuiper belt objects can be divided in three classes according to
their orbits: classical, resonant and scattered. The first model we have 
studied is based on the idea that all the Kuiper Belt mass is
concentrated in the main resonant orbits. The second considered model
assumes that all the Kuiper Belt mass is uniformly concentrated
in the ecliptic disk, likewise most of the classical objects.
The third studied model is based on the assumption that all the Kuiper
Belt mass is non-uniformly concentrated in the ecliptic disk as suggested for 
the first time by Boss and Peale \cite{Boss}. Finally, the last model we 
have considered explores the idea that all the Kuiper Belt mass is 
uniformly concentrated in a torus, encompassing also the objects
outside the ecliptic. Each of these models encompass some of the features of the Kuiper
Belt.

Actually,our approach is the opposite of the one adopted in Ref. \cite{Diego}, where the Kuiper Belt 
structure and mass were chosen to fit the Pioneer anomaly. As already mentioned this 
implies that the mass of Belt is about $1.97$ Earth masses, far too great to account for the observations, as 
already mentioned ($0.3$ Earth mass is the figure of merit \cite{Backman}), and that is starts at about $20$ AU, 
which has little support on the facts.

In what follows, we consider in detail the gravitational forces generated by four different models of the Kuiper Belt. We 
point out that we consider here the toroidal model that has not been previously discussed in the literature and 
also consider for the first time the gravitational force away from the ecliptic for all models that we discuss. 

\subsubsection{The two-ring models}

\begin{figure}[h]
\begin{center}
\begin{tabular}{cccc}
\multicolumn{1}{c}{}&\\
\epsfxsize=12.3 cm
\epsffile{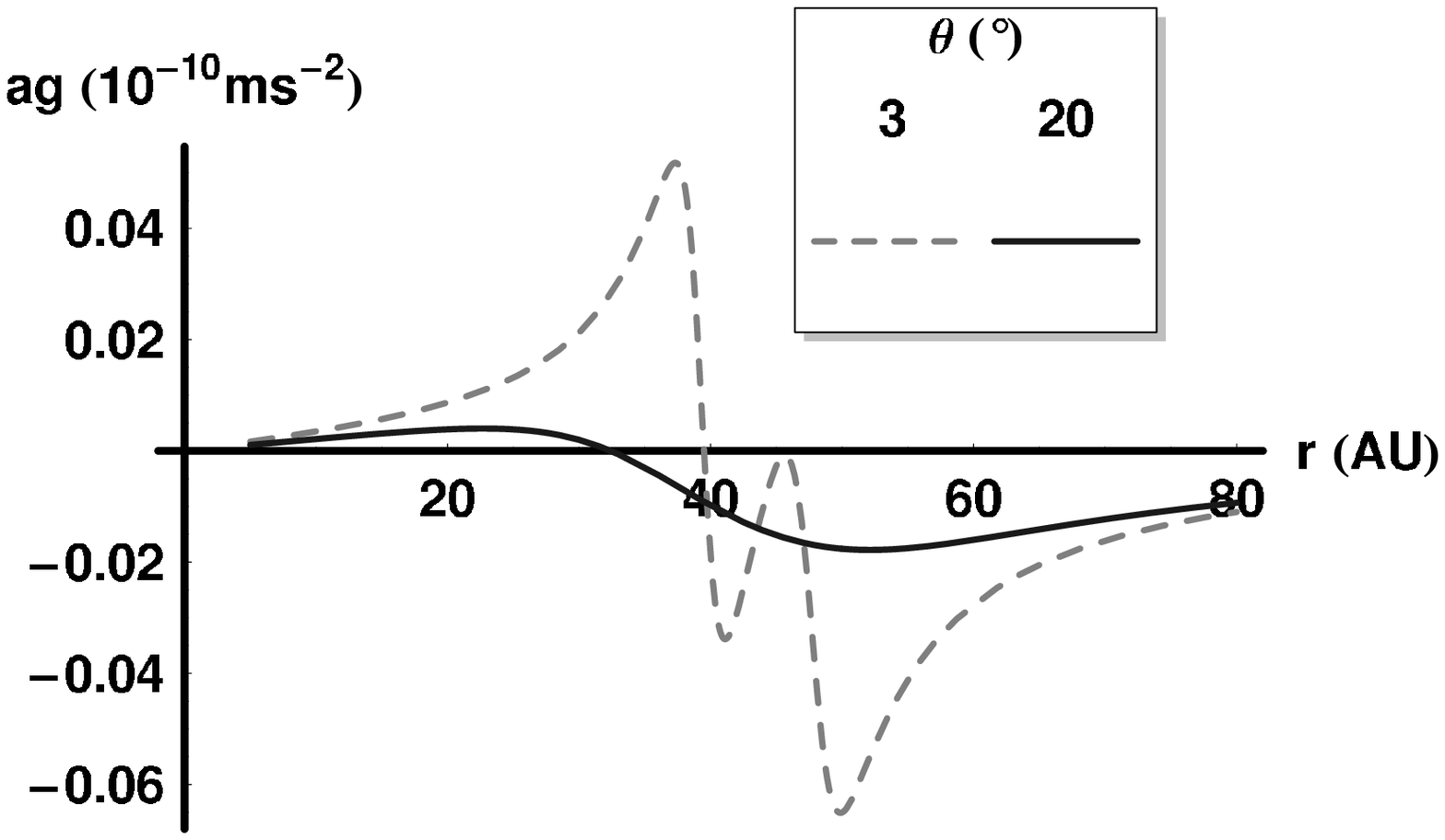}  \\
\mbox{(a) The dash line represents $\theta=3^{o}$ and the full line $\theta=20^{o}$. } \\ 
\\
\\
\epsfxsize=12.25 cm
	\epsffile{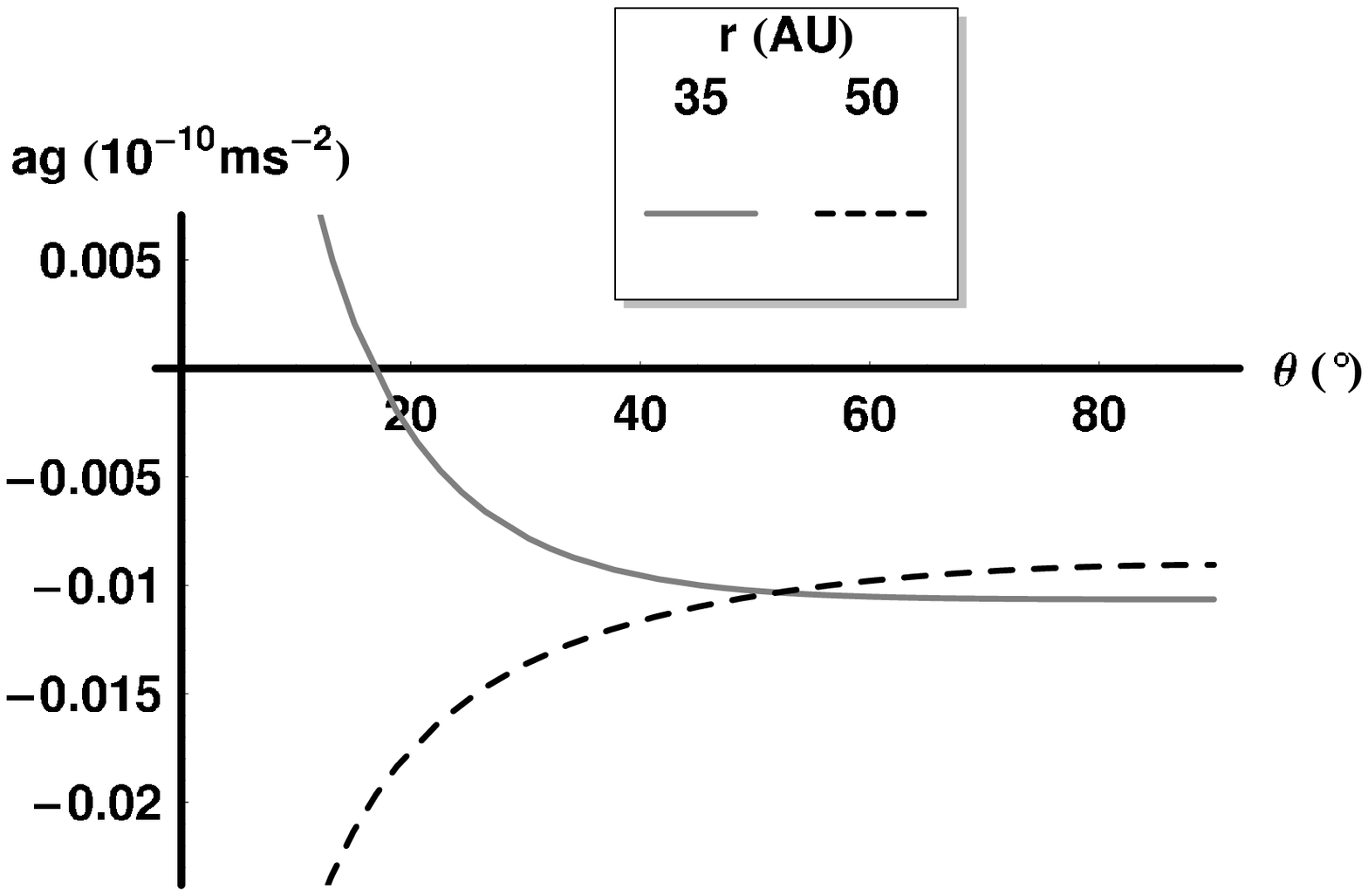}  \\
\mbox{(b) The full line represents $r=35 AU$ and the dash line $r=50 AU$.}\\
\end{tabular}
\end{center}
\caption{Variation of the gravitational acceleration, 
due to the two-ring model of the Kuiper Belt in terms of the distance and the 
solar ecliptic latitude angle. }
\label{fig:anel}
\end{figure}

We consider first the two-ring model which consists of two thin rings lying on the ecliptic and with 
radius $R_{1}=39.4$ AU (resonance 3:2) and $R_{2}=47.8$ AU (resonance 2:1). 
The radial acceleration yielded by this model at a distance r from the Sun is given by:
\begin{equation}
a_{rg}= - \frac{G M_{KB}}{2 \pi (R_{1}+R_{2})}\int_{0}^{2 \pi}  \Big( \sum_{i=1}^{2} R_{i}  g_{rg}(\phi_{m}) \Big) \, d\phi_{m} 
\end{equation}  
with
\begin{equation}
 g_{rg}(\phi_{m})= \frac{r-R_{i} cos \theta cos (\phi_{m}-\phi)}{(r^{2}+R_{i}^{2}-2 r R_{i} cos \theta cos (\phi_{m}-\phi) )^\frac{3}{2}}
\end{equation}  
where $\phi_{m}$ is the solar ecliptic longitude of the element of mass distribution, 
$M_{KB}$ is the total mass of the Kuiper Belt,
$r$, $\theta$ and $\phi$ are the spherical coordinates of the solar ecliptic coordinate system of the probe.
The radial acceleration produced by the two-ring model has been obtained numerically and shown in Figure \ref{fig:anel}.


\subsubsection{The uniform disk model}

\begin{figure}[h]
\begin{center}
\begin{tabular}{cccc}
\multicolumn{1}{c}{}&\\
\epsfxsize=12.3 cm
\epsffile{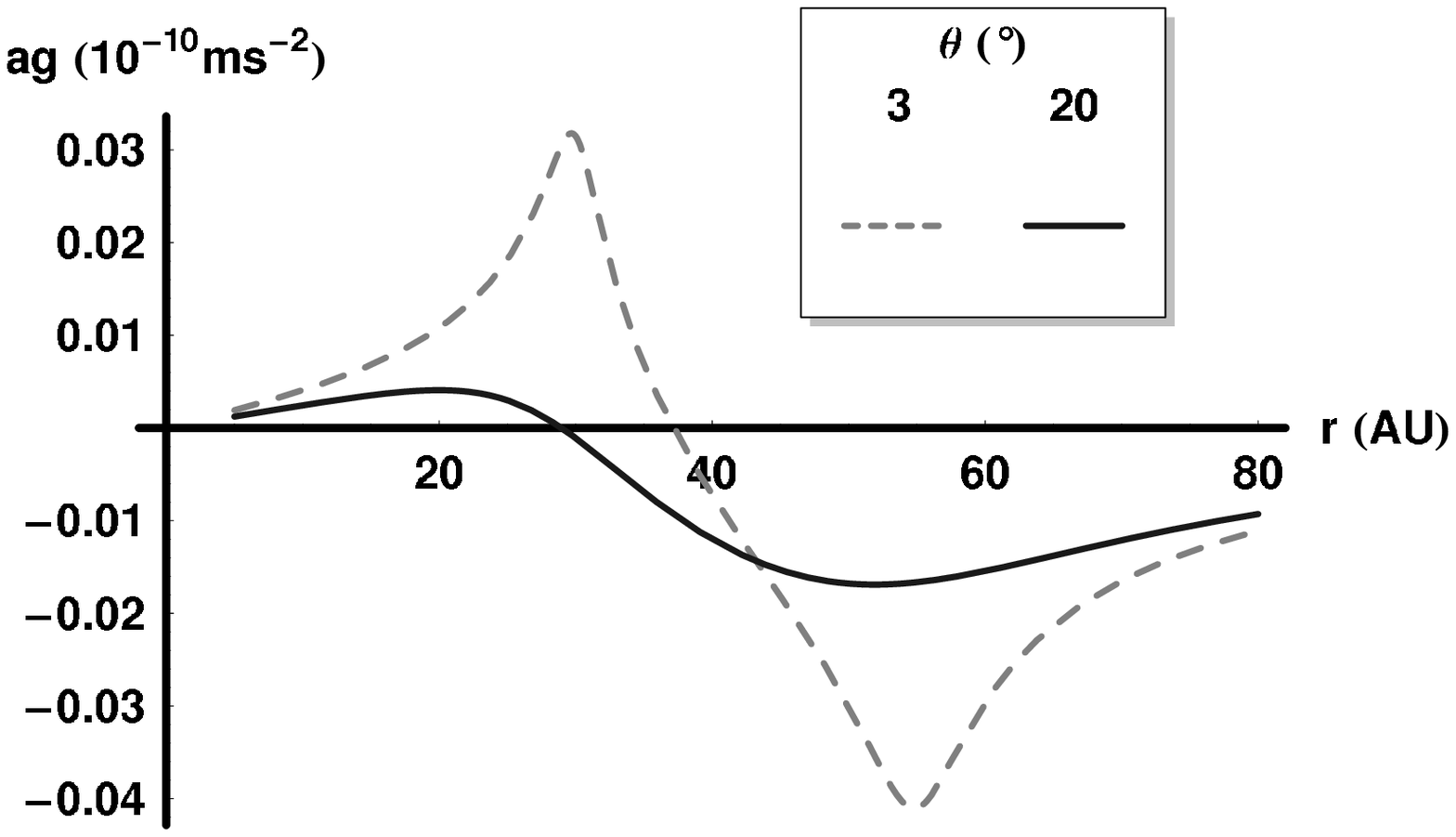}  \\
\mbox{(a) The dash line represents $\theta=3^{o}$ and the full line $\theta=20^{o}$. } \\ 
\\
\\
\epsfxsize=12.25 cm
	\epsffile{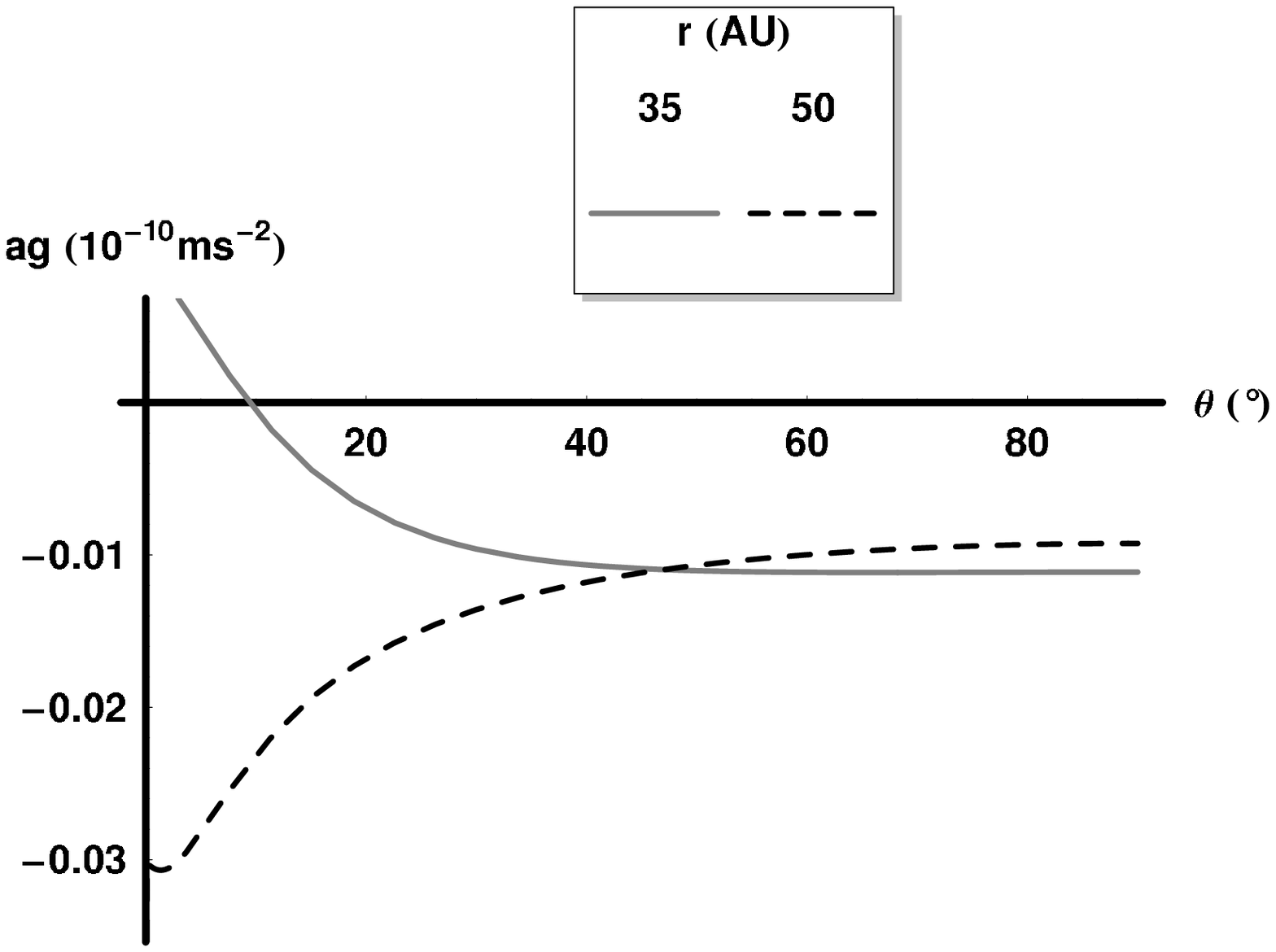}  \\
\mbox{(b) The full line represents $r=35 AU$ and the dash line $r=50 AU$. }\\
\end{tabular}
\end{center}
\caption{Variation of the gravitational acceleration, 
generated by the uniform disk model of the Kuiper Belt in terms of the distance and the 
solar ecliptic latitude angle.}
\label{fig:disco}
\end{figure}

Next we consider the uniform disk model, which consists of an uniform hollow thin disk 
lying on the ecliptic, within distances $R_{min}=30$ AU and $R_{max}=55$ AU. 
The radial acceleration at a distance $r$ from the Sun is given by:
\begin{equation}  
a_{ud}= - \frac{G  M_{KB}}{R_{max}^{2}-R_{min}^{2}} \int_{0}^{2 \pi}  
\int_{R_{min}}^{R_{max}} g_{ud }(r_{m},\phi_{m}) \, dr_{m}  \, d\phi_{m}
\end{equation}  
with
\begin{equation}  
g_{ud}(r_{m},\phi_{m})=\frac{ r_{m} (r-r_{m} cos \theta cos (\phi_{m}-\phi)) }{(r^{2}+r_{m}^{2}-2 r r_{m} 
cos \theta cos (\phi_{m}-\phi) )^\frac{3}{2}}
\end{equation}  
where $r_{m}$ is the solar ecliptic radius of the element of mass distribution. 
The radial acceleration produced by the uniform disk model has been obtained numerically and 
exhibited in Figure \ref{fig:disco}.


\subsubsection{The non-uniform disk model}

We consider now the non-uniform thin disk model discussed for the first time by Boss and 
Peale \cite{Boss} and use $R_{min}=30$ AU and $R_{max}=100$ AU. We have generalized the discussion of these authors 
to any point in three-dimensional space, introducing the term $cos \theta$ in their formula \cite{Boss}.

For this model of the Kuiper Belt the radial acceleration at a distance $r$ from the Sun is given by: 
\begin{equation}  
a_{nud}= -\frac{0.1 G  M_{KB}}{ R_{max}^{2}-R_{min}^{2}} \int_{R_{min}}^{R_{max}}  \int_{0}^{2 \pi} 
g_{nud}(r_{m},\phi_{m})  \, d\phi_{m} \, dr_{m}
\end{equation}   
with  
\begin{equation}  
g_{nud}(r_{m},\phi_{m})= f(r_{m})\frac{r_{m} r -r_{m}^{2} cos \theta cos(\phi_{m}-\phi) }{(r^{2}+r_{m}^{2}-2 r_{m} r 
cos \theta cos(\phi_{m}-\phi))^{3/2}}
\end{equation}  
and
$f(r_{m})=\frac{(r_{m}-R_{min})^2}{AU^2} e^{\frac{-0.2 (r_{m}-R_{min})}{AU}} .$

\begin{figure}[h]
\begin{center}
\begin{tabular}{cccc}
\multicolumn{1}{c}{}&\\
\epsfxsize=12.3 cm
\epsffile{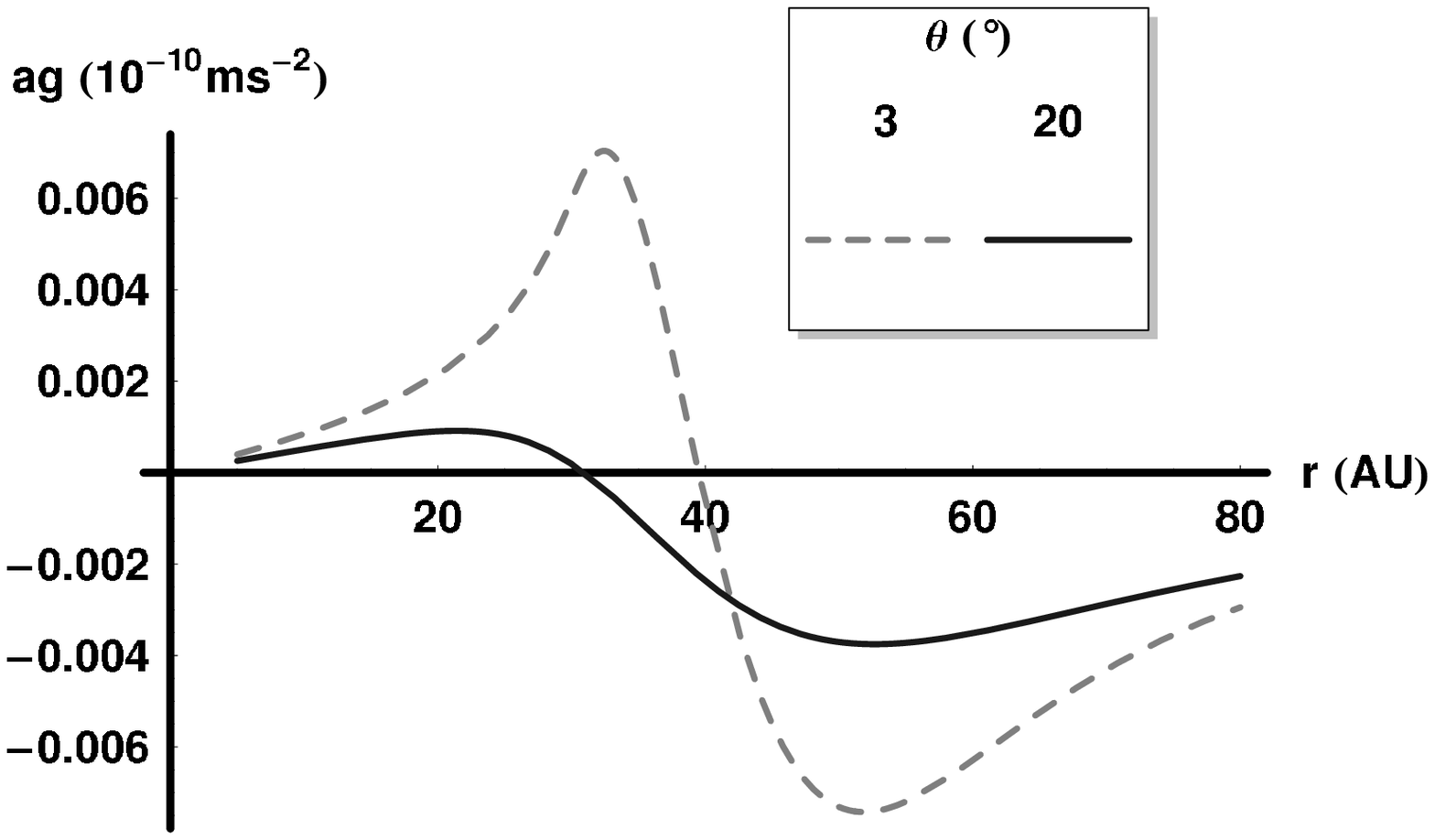}  \\
\mbox{(a) The dash line represents $\theta=3^{o}$ and the full line $\theta=20^{o}$. } \\ 
\\
\\
\epsfxsize=12.3 cm
	\epsffile{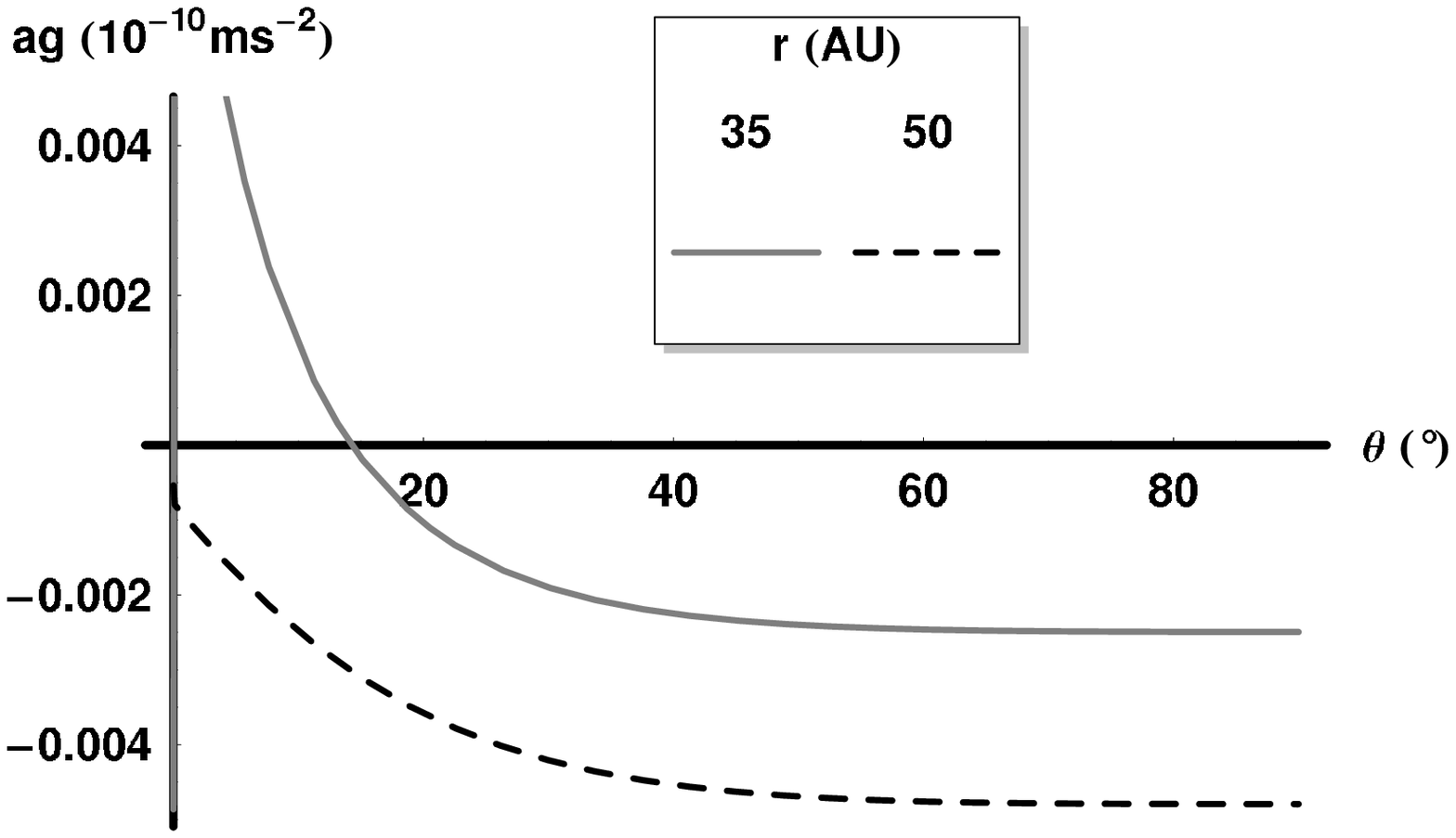}  \\
\mbox{(b) The full line represents $r=35 AU$ and the dash line $r=50 AU$. }\\
\end{tabular}
\end{center}
\caption{Variation of the gravitational acceleration, 
produced by the non-uniform disk model of the Kuiper Belt in terms of the distance and the 
solar ecliptic latitude angle.}
\label{fig:disconunif}
\end{figure}

The radial acceleration produced by the non-uniform disk model has been obtained numerically and is shown in Figure \ref{fig:disconunif}.
Analytic techniques to study the gravitational force for this model have been discussed in Ref. \cite{Nieto1}.

\subsubsection{The toroidal mass distribution model}

Finally, we consider
the torus model, i.e. a toroidal mass distribution centered over the ecliptic with central 
radius $R_{c}=42.5$ AU and thickness $R_{t}=12.5$ AU.
The radial acceleration at a distance $r$ from the Sun produced by the Kuiper Belt is: 
\begin{equation}  
a_{tor}=-\frac{G M_{KB}}{2 \pi^{2} R_{t}^{2}} \int_{0}^{R_{t}} \int_{0}^{2\pi}  \int_{0}^{2\pi} 
g_{tor}(r_{m},\beta_{m},\phi_{m})  \, d\phi_{m} \, d\beta_{m} \, dr_{m} 
\end{equation}  
where,
\begin{eqnarray}  
g_{tor}(r_{m},\beta_{m},\phi_{m}) = \frac{-(r_{m} R_{c} + r_{m}^{2} cos \beta_{m} ) cos \theta cos ( \phi-\phi_{m} ) + r_{m} r - r_{m}^{2} 
sin \beta_{m} sin \theta}{ ( h(r_{m},\beta_{m},\phi_{m})  + r_{m}^{2} + r^{2} + R_{c}^{2} )^{\frac{3}{2}}}
\end{eqnarray}  
and
\begin{eqnarray}  
h(r_{m},\beta_{m},\phi_{m})=2 r_{m} cos \beta_{m} R_{c}-2 r_{m} r sin\beta_{m} sin \theta -(2 r r_{m} cos \beta_{m} +2 r R_{c}) cos\theta 
cos(\phi-\phi_{m})
\end{eqnarray} 
with $\beta_{m}$ being the angle from the ecliptic to the element of mass distribution with origin at $R_{c}$. 

\begin{figure}[h]
\begin{center}
\begin{tabular}{cccc}
\multicolumn{1}{c}{}&\\
\epsfxsize=12.3 cm
\epsffile{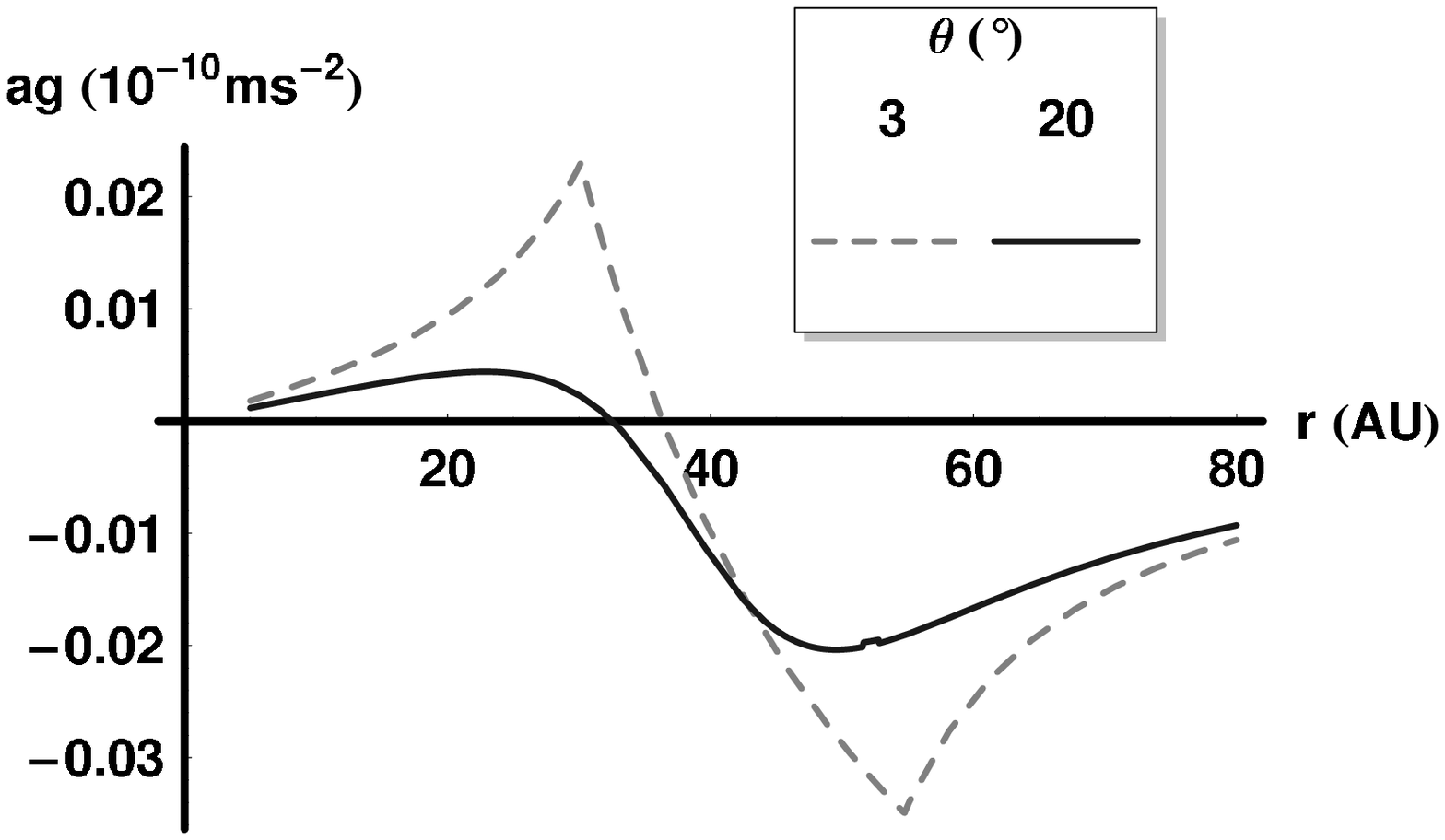}  \\
\mbox{(a) The dash line represents $\theta=3^{o}$ and the full line $\theta=20^{o}$. } \\ 
\\
\\
\epsfxsize=12.3 cm
	\epsffile{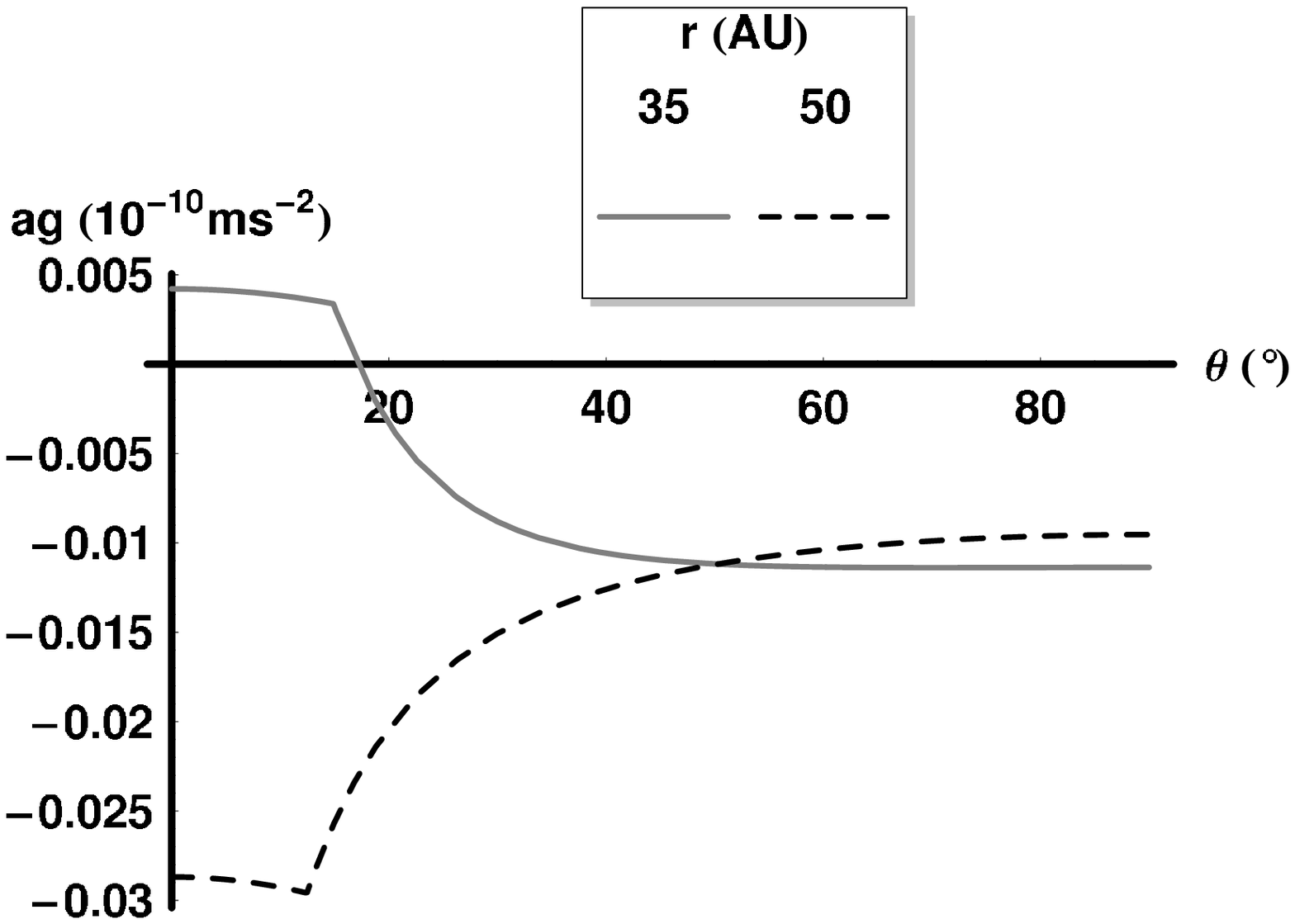}  \\
\mbox{(b) The full line represents $r=35 AU$ and the dash line $r=50 AU$. }\\
\end{tabular}
\end{center}
\caption{Variation of the gravitational acceleration, 
generated by the uniform torus model of the Kuiper Belt in terms of the distance and the 
solar ecliptic latitude angle.}
\label{fig:toro}
\end{figure}

The radial acceleration produced by the uniform torus model has been computed numerically and depicted in Figure \ref{fig:toro}.


\begin{figure}[h]
\begin{center}
\begin{tabular}{cccc}
\multicolumn{1}{c}{}&\\
\epsfxsize=12.3 cm
\epsffile{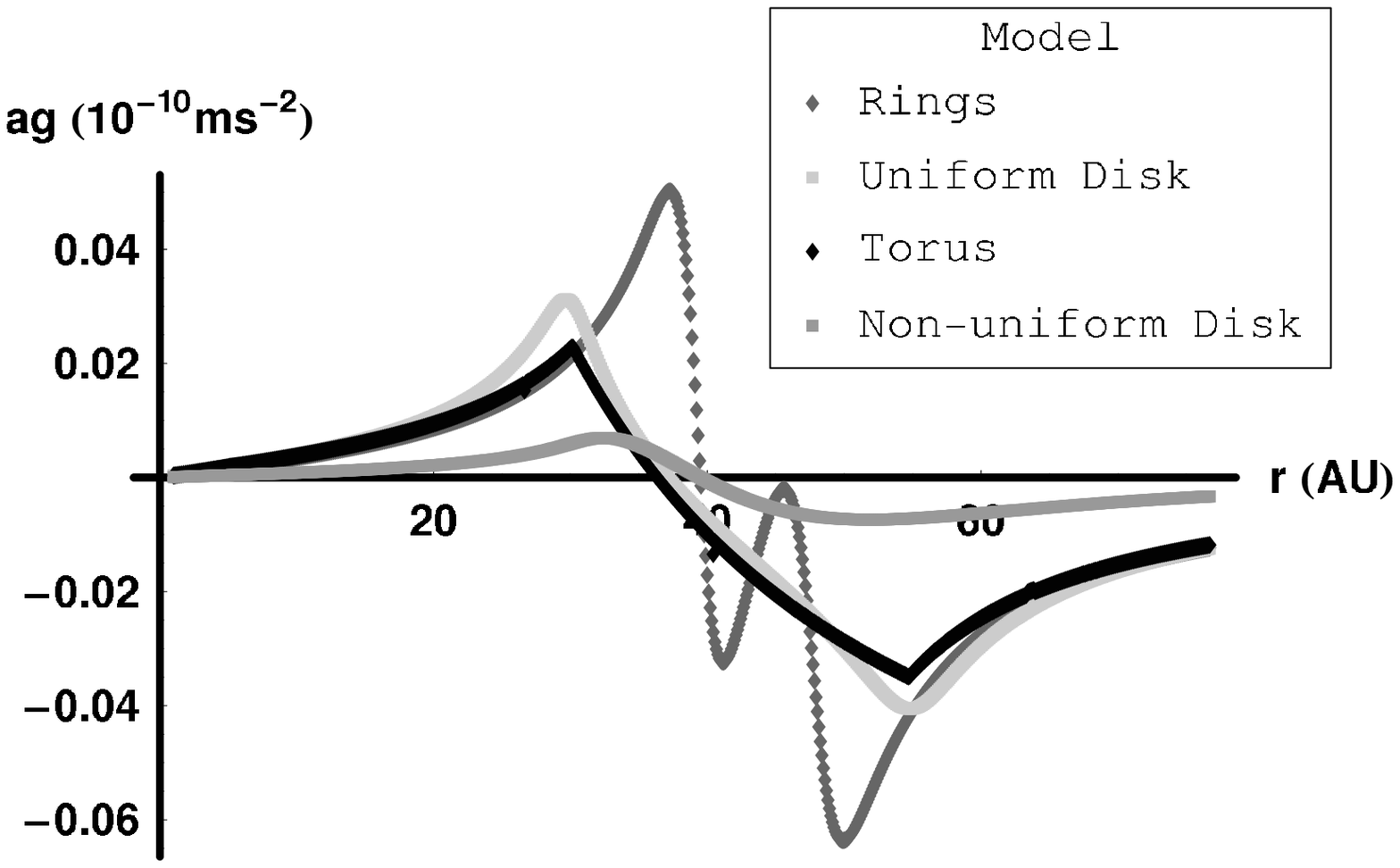} 
\end{tabular}
\end{center}
\caption{Gravitational acceleration acting on the Pioneer 10 for all the models.
The lighter grey represents the uniform disk, the medium grey the non-uniform disk, 
the dark grey the two-ring model and the black the torus model}
\label{fig:pioneer}
\end{figure}

Through the use of these expressions for the radial accelerations, we have computed
the gravitational acceleration due to the various models of Kuiper Belt as probed by the Pioneer 10
spacecraft. For that we have used the Pioneer data of Ref. \cite{Helioweb}. Our results are shown in Figure
\ref{fig:pioneer}.


\begin{figure}[h]
\begin{center}
\begin{tabular}{cccc}
\multicolumn{1}{c}{}&\\
\epsfxsize=12.3 cm
\epsffile{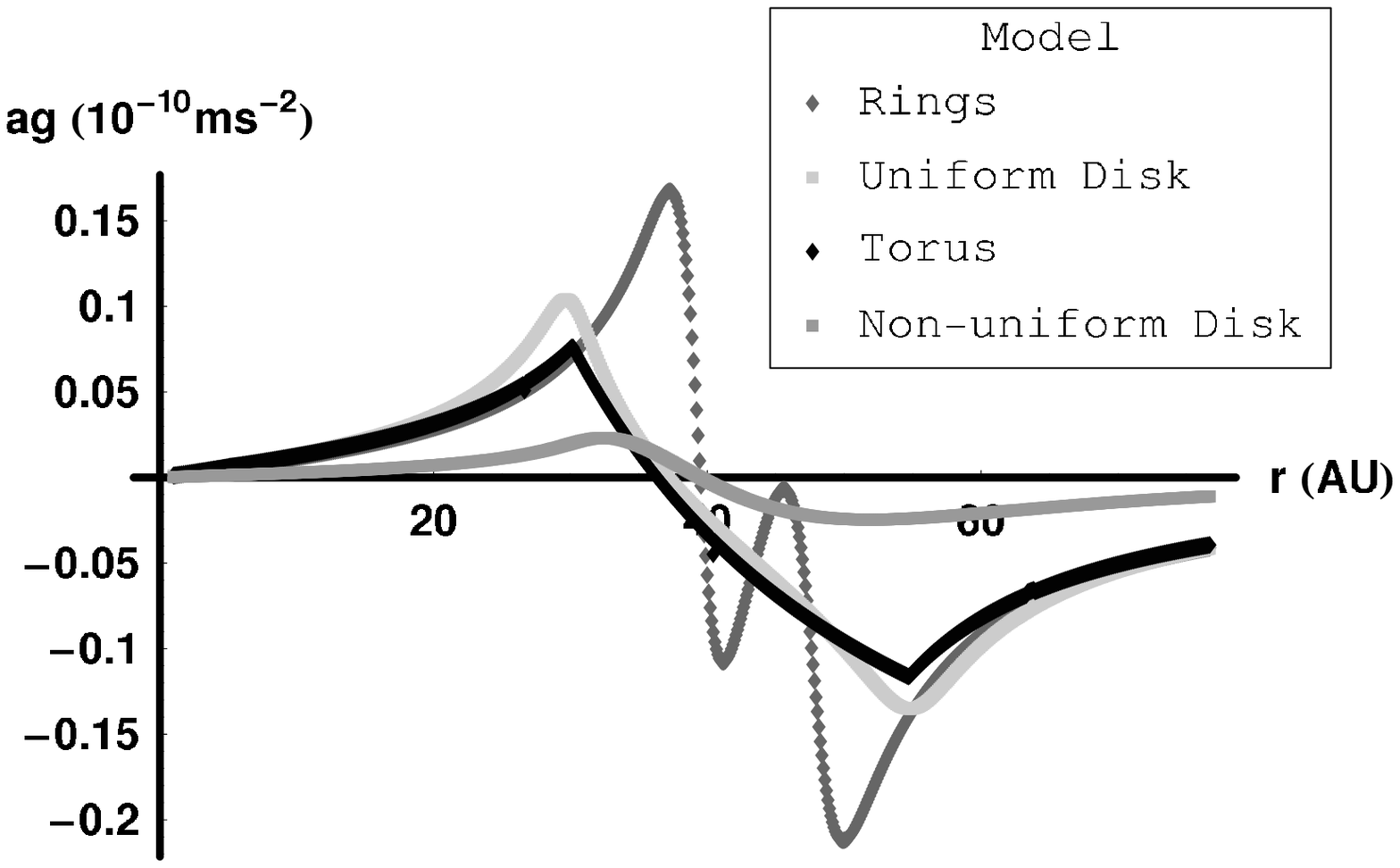}
\end{tabular}
\end{center}
\caption{Gravitational acceleration acting on the Pioneer 10 spacecraft 
for all the models with a total mass for Kuiper Belt of 1 Earth mass.
The lighter grey represents the uniform disk, the medium grey the non-uniform disk, 
the dark grey the two-ring model and the black the torus model}
\label{fig:pioneer1}
\end{figure}

For all the models that we have studied, the gravitational acceleration changes
sign between 35 and 40 AU. Until about 35 AU, the acceleration is positive for all the models, 
while after 40 AU, it turns negative for all models.
The closest model to yield a constant acceleration, a distinct observational feature of the 
Pioneer anomaly, is the
non-uniform disk model, however the found magnitude is much smaller than the
anomalous acceleration. The model which gives rise to the greatest acceleration is the two-ring model.
We can see that, for all the cases, the maximum acceleration that one obtains 
is $0.064 \times 10^{-10}~m s^{-2}$, which is
$0.73 \%$ of the observed anomalous acceleration. Even if one considers the total mass of the
Kuiper Belt as being 1 Earth mass (Figure \ref{fig:pioneer1}), the
maximum value for the acceleration that one obtains is $0.212
\times 10^{-10}~m s^{-2}$, which is $2.43 \%$ of the anomalous
acceleration. Hence, our results clearly indicate that the
gravitational acceleration due to the Kuiper Belt cannot be the
cause of the Pioneer anomaly. 


Notice that the above expressions for the acceleration have singularities in 
the situation where the craft lies inside the mass distribution.
The uniform and non-uniform disk models 
have singularities for $\theta \sim 0$. 
The torus model has singularities at $\theta=3^{o}$ for $r$ inside the torus ($30$ AU $\le r \le 55$ AU), at 
$r=35$ AU for $0^{o} \le \theta \le 15.75 ^{o}$ and at $r=50$ AU for $11.25^{o} \le \theta \le 13.5^{o}$.

\section{Pioneer anomaly and drag forces}

\subsection{The effect of drag forces on Pioneer 10}

Having established that the gravitational acceleration of the various models of the Kuiper Belt 
cannot account for the Pioneer anomaly, we study now to which extent drag forces might 
have affected the motion of the Pioneer spacecraft. As we shall see the inferred dust distribution required 
to explain the anomaly is many orders of magnitude greater than the interplanetary and interstellar dust densities 
as well as considerably greater than ones that can be inferred from the Kuiper Belt that we considered in 
the previous section.  

In oder to perform this analysis we examine the deceleration caused by drag forces on the spacecraft due to 
the interplanetary medium. This force can be modeled as \cite{Nieto2} :
\begin{equation}
a_{s}(r)=-\frac{K_{s} \rho (r) v_{s}(r)^{2} A_{s} }{m_{s}}~,
\end{equation}
where $\rho (r)$ is an in situ density, presumably the one of the interplanetary medium,
$K_{s}$ an $O(1)$ coefficient associated with reflection $(K_{s} =
2)$, absorption $(K_{s} = 1)$ and transmission $(K_{s} = 0)$ of
particles hitting the craft, $v_{s}$ is the velocity of the craft
with respect to the medium, $A_{s}$ the effective cross-sectional
area of the craft, and $m_{s}$ its mass. For the Pioneer spacecraft
$v_{s}=11.6-12.2~km/s$, $m_{s}=241 kg$ (after consumption of half of
the fuel) and $A_{s}=5.9~m^{2}$. Assuming that the Pioneer anomaly
can be regarded as an in situ measurement of the acceleration, that
is $a_{s}(r)=a_{p}$, the resultant density is 
$2.48 \times 10^{-19} ~g cm^{-3}~$. 
We can compare the density profile obtained
with estimated values of the interplanetary dust density
\cite{Mann,Kelsall}
\begin{equation}
\rho_{IPD} \ge  10^{-24} ~g cm^{-3}~,
\end{equation}
and the interstellar dust density \cite{Mann}
\begin{equation}
\rho_{ISD} \le 3 \times 10^{-26} ~g cm^{-3}~.
\end{equation}

Density profiles of the form
\begin{equation}
\rho_{n}(r) = \rho_{0 n}(r) \left(\frac{r_{0}}{r}\right)^n~,
\end{equation}
for $\rho_{0 n}$ and $r_{0}$ constant, have also been suggested for
comparison \cite{Nieto2}. For $n=0$, the density is uniform and its
estimated value is \cite{Nieto2}
\begin{equation}
\rho_{0} = \rho_{0 0} \equiv \rho_{p} \le 3 \times 10^{-19} ~g cm^{-3}~, 
\end{equation}
which corresponds to the one required to explain the anomaly, $\rho_{p}$.

\begin{figure}[h]
\begin{center}
\begin{tabular}{cccc}
\multicolumn{1}{c}{}&\\
\epsfxsize=12.3 cm
\epsffile{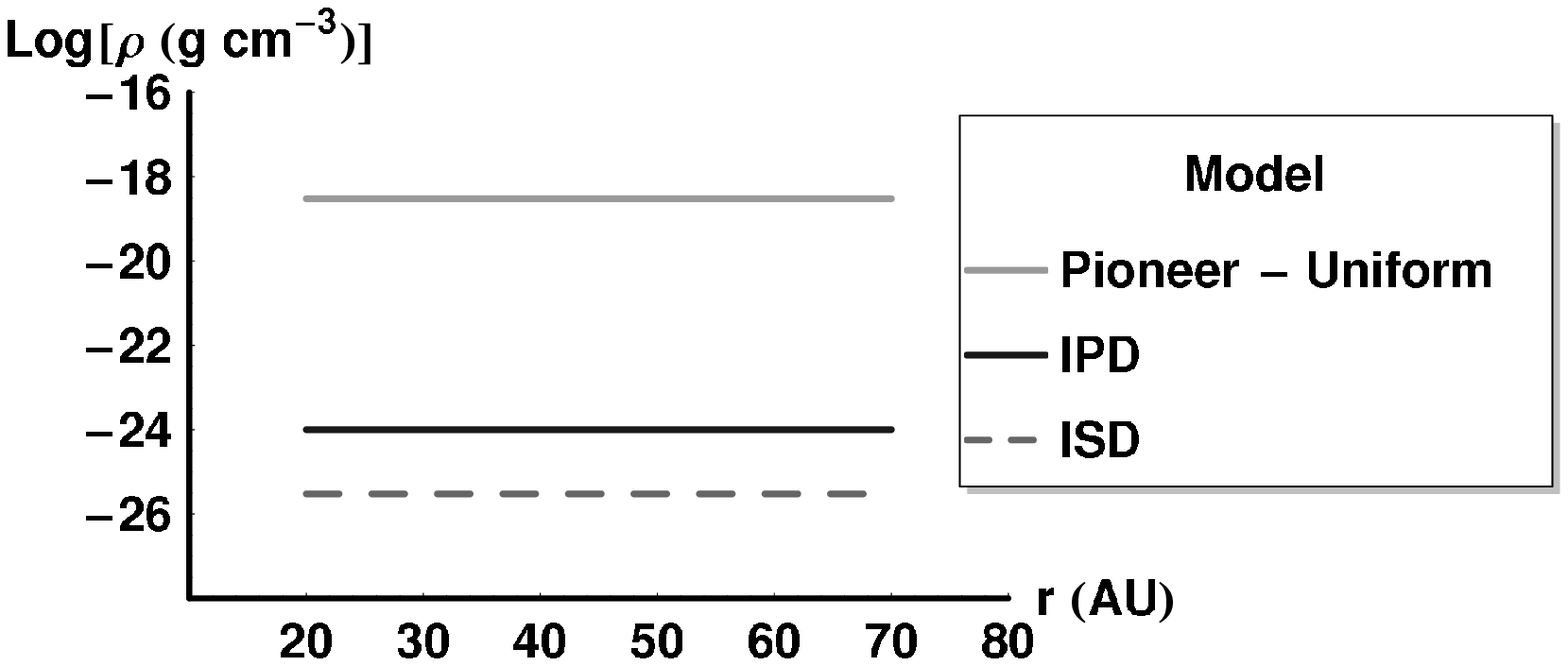}
\end{tabular}
\end{center}
\caption{Density profiles of ISD (grey dash line), IPD (black full line) and the uniform density obtained
using the Pioneer anomalous acceleration as an in situ measurement (grey full line).}
\label{fig:denstodos}
\end{figure}

The comparison between the density obtained assuming the Pioneer
anomalous acceleration as an in situ measurement, the interstellar
and interplanetary dust densities and the uniform density suggested
in Ref. \cite{Nieto2} is depicted in Figure \ref{fig:denstodos}. It is clear that 
a mismatch of several orders of magnitude exist between the estimated 
amount of interplanetary and interstellar dust and the 
one required to explain the Pioneer anomaly.


\subsection{Drag forces and the Kuiper Belt}

In the previous section we have considered several models for the Kuiper Belt in order 
to estimate the resultant gravitational force acting on a spacecraft, we consider now the density of 
each of the analyzed models in order to compare with the density necessary to explain the anomaly due to 
drag forces.  The total mass of Kuiper Belt is taken to $0.3$ Earth masses.

For the two-ring model, the density was given by $M_{KB}/2 \pi (R_{1}+R_{2})$. We consider now the
following three-dimensional estimate:
\begin{equation}
\rho_{rg}=\frac{M_{KB}}{2 \pi \delta r \delta z (R_{1}+R_{2})} ~,
\end{equation}
with $\delta z =16$ AU and $\delta r=\delta z /2 = 8$ AU. The value of $\delta z$ is estimated
considering the double of the average of height above the ecliptic of the Kuiper Belt Objects.

For the uniform disk model, the considered density was given by
$M_{KB}/ \pi  (R_{2}^2-R_{1}^2)$. We consider instead, the density in three dimensions as
\begin{equation}
\rho_{ud}=\frac{M_{KB}}{\pi \delta z (R_{2}^2-R_{1}^2)} ~.
\end{equation}

On its turn, the density for the non-uniform disk model was taken to be
$0.1~M_{KB} f(r_{m})/ \pi  (R_{2}^2-R_{1}^2)$.
In three dimensions we consider the following density
\begin{equation}
\rho_{nud}=\frac{0.1 M_{KB} f(r_{m})}{\pi \delta z
(R_{2}^2-R_{1}^2)}~.
\end{equation}

Finally, for the uniform torus model in three dimensions is given by:
\begin{equation}
\rho_{t}=\frac{M_{KB}}{2 \pi^{2} R_{t}^{2} R_{c}}~.
\end{equation}

\begin{figure}[h]
\begin{center}
\begin{tabular}{cccc}
\multicolumn{1}{c}{}&\\
\epsfxsize=15.3 cm
\epsffile{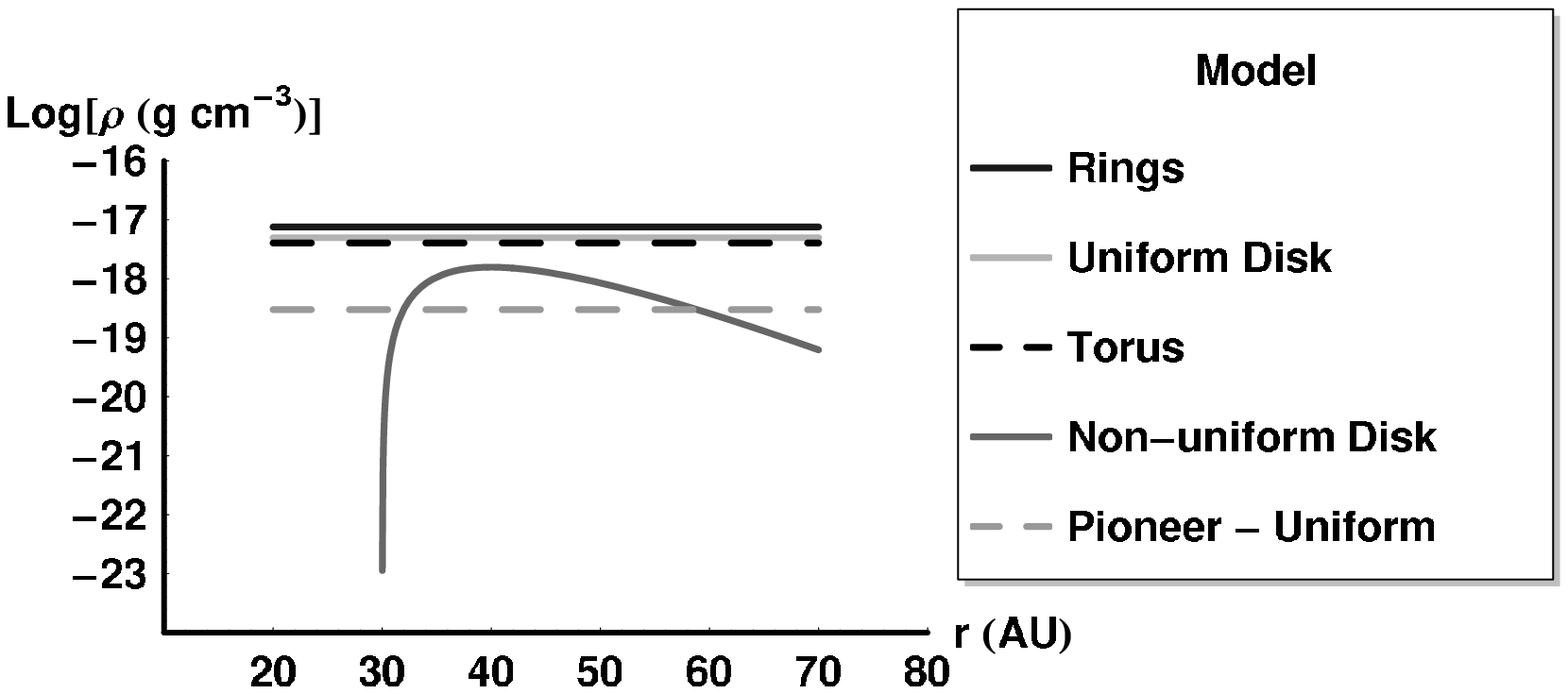}
\end{tabular}
\end{center}
\caption{Density profiles of the Kuiper Belt models with $0.3$ Earth mass 
and the density profile obtained
using the Pioneer anomalous acceleration as an in situ measurement.
The two-ring model is represented by the black full line, the uniform disk by the light grey full line, 
the torus model by the black dashed line, the non-uniform disk by the dark grey full line and the the density obtained
from Pioneer anomalous acceleration by the light grey dashed line.}
\label{fig:dens}
\end{figure}

Figure \ref{fig:dens} exhibits the density profiles of the considered Kuiper Belt
models and the density profile obtained assuming the Pioneer
anomalous acceleration as an in situ measurement. Our results can be summarized as follows: 
the densities of the two-ring model, the uniform disk and the toroidal model are uniform, but are about 
two orders of magnitude greater than the one necessary to explain the anomaly, $\rho_{p}$; 
for the non-uniform disk, we find that the computed density is greater than $\rho_{p}$ between 
$35$ AU to about $62$ AU, and smaller elsewhere.

\section{Conclusions}

In this work we have shown that the Pioneer anomaly cannot be
explained by the gravitational acceleration of the various Kuiper
Belt models considered (two-ring, uniform and non-uniform disk, and
toroidal mass distribution). In none of the studied cases a constant
acceleration was found and the order of magnitude of the obtained
accelerations is at best about a few percent of the observed
anomalous acceleration.

We have also verified that if one assumes that the Pioneer density is due to drag forces acting on the spacecraft 
the required density is many orders of magnitude greater than the accepted for the interplanetary and
interstellar dust densities. Furthermore, we have compared these densities 
with the ones corresponding the various Kuiper Belt mass distribution models and found  
that the the densities of the two-ring model, the uniform disk and the toroidal model are about two  
orders of magnitude greater than the one necessary to account for the anomaly. This means that 
for the two-ring model, the uniform disk and the toroidal model the anomalous 
acceleration can be explained by drag forces only if about less than one percent of mass distribution 
contributes to the drag. For the non-uniform disk, explaining the Pioneer anomaly via a drag force can be 
most likely excluded as only at $35$ AU and about $62$ AU, the computed density matches $\rho_{p}$.

These conclusions provide further support to the idea that only through a mission (dedicated or not)
\cite{Bertolami3,Anderson2,Bertolami4,Dittus,Izzo,Bertolami5} 
this mysterious anomalous acceleration can be confirmed and more thoroughly 
scrutinized. It is quite pleasing that this possibility is being seriously considered by ESA in its Cosmic 
Vision programme for the 2015 - 2025 
period \cite{ESA}.

\vskip 2cm

{\bf Acknowledgments}

\vskip 0.5cm

\noindent
It is a pleasure to thank Clovis de Matos, Klaus Scherer and Slava Turyshev 
for helpful comments and discussions.

\newpage

\vfill

\end{document}